\documentclass[twocolumn,aps,pra,10pt,showpacs]{revtex4-1}
\usepackage{bm}
\usepackage{amsfonts}
\usepackage{amssymb}
\usepackage{amsmath}
\usepackage{graphicx}
\begin{document}

\title{Twisted Localized Solutions of the Dirac Equation:\\Hopfion-like States of Relativistic Electrons}
\author{Iwo Bialynicki-Birula}\email{birula@cft.edu.pl}
\affiliation{Center for Theoretical Physics, Polish Academy of Sciences\\
Aleja Lotnik\'ow 32/46, 02-668 Warsaw, Poland}
\author{Zofia Bialynicka-Birula}
\affiliation{Institute of Physics, Polish Academy of Sciences\\
Aleja Lotnik\'ow 32/46, 02-668 Warsaw, Poland}

\begin{abstract}

All known solutions of the Dirac equation describing states of electrons endowed with angular momentum are very far from our notion of the electron as a spinning charged bullet because they are not localized in the direction of propagation. We present here normalizable analytic exact solutions, eigenstates of the total angular momentum component $M_z$, that come very close to this notion. These new solutions of the Dirac equation have also intricate topological properties similar to the hopfion solutions of the Maxwell equations.
\end{abstract}
\maketitle

\section{Introduction}

The purpose of this work is to present a family of exact solutions of the Dirac equation that have properties similar to the solutions of Maxwell equations --- called hopfions --- describing structured light. Studies of structured light, both theoretical and experimental, have become a well developed field of research described in hundreds of publications \cite{rub}. A significant part of this field, marked by a sophisticated mathematical component is devoted to knotted light \cite{abt}. This subject started with the discovery by Ra$\tilde{\rm n}$ada \cite{ran} that a certain compact solution of Maxwell equations represents a physical model of the Hopf fibration. This solution, named now the hopfion, was known to Synge \cite{syn} but he did not discover its topological content.

In recent years, structured beams have also become an important experimental and theoretical field of research for the electrons. The scale of interest in this topic is illustrated by 334 references (most of them fairly new) in the latest review paper \cite{bliokh}.

In this work we contribute to the theoretical understanding of the structured electron states by exhibiting analytic solutions of the Dirac equation with intricate topological properties. These solutions describe the states of a relativistic electron in free space endowed with angular momentum. In contrast to the known beam-like solutions (Bessel, Laguerre-Gauss, and exponential beams, cf.~\cite{bliokh,lloyd,karimi,bb0}), they are fully {\em localized} in the three-dimensional space; they decrease exponentially in {\em all} directions. It seems to us that such localized wave packets are more suitable to describe the experiments with relativistic electrons than the wave functions with an unlimited extension in the direction of propagation.

Our solutions will be generated from a single complex function $f_{\rm KG}(\bm r,t)$ satisfying the Klein-Gordon (KG) equation. By the $l$-th differentiation with respect to $x-iy$ we generate from $f_{\rm KG}(\bm r,t)$ the solutions $f_l(\bm r,t)$ of the Klein-Gordon equation with the orbital angular momentum in the $z$ direction equal to $\hbar l$. Finally, using the procedure described in \cite{bb0}, denoted as KG$\to$D, we generate from the scalar functions $f_l(\bm r,t)$ the solutions of the Dirac equation which have topological properties quite similar to those of the electromagnetic hopfions.

\section{Localized solutions of the Dirac equation}

The Dirac equation is a set of four coupled equations. The task of simultaneously satisfying all equations is quite cumbersome. However, by starting from a single function $f_{\rm KG}(\bm r,t)$ satisfying the KG equation we may reduce this task to straightforward differentiations \cite{bb0}. We choose here the basic generating function $f_{\rm KG}(\bm r,t)$ as a close analog of the Synge solution of the d'Alembert equation describing a state of the photon. In the present case we have the exponential wave packet in momentum space, describing a state of en electron (not a positron), built from the plane waves with positive energy $(c=1,\hbar=1)$,
\begin{align}\label{kg}
f_{\rm KG}({\bm r},t)=\frac{1}{4\pi}\int \frac{d^3p}{E_p}e^{-(a+it)E_p}e^{i{\bm p}\cdot{\bm r}},
\end{align}
where $E_p=\sqrt{m^2+{\bm p}^2}$ and $a$ measures the spatial size of the wave packet. The integration over $\bm p$ leads to,
\begin{align}\label{kg1}
f_{\rm KG}({\bm r},t)=\frac{mK_1(ms)}{s},
\end{align}
where $s=\sqrt{x^2+y^2+z^2+(a+it)^2}$ and $K_1(ms)$ is the Macdonald function. In the massless limit one obtains the solution of the d'Alembert equation used by Synge,
\begin{align}\label{synge}
\lim_{m\to 0} f_{\rm KG}({\bm r},t)=\frac{1}{s^2}.
\end{align}
Since derivatives of a solution of the KG equation are also solutions, we can extend $f_{\rm KG}$ from $l=0$ to an arbitrary value of the orbital angular momentum,
\begin{align}\label{arb}
f_l({\bm r},t)&=(-1/m)^l (\partial_x+i\partial_y)^lf_{\rm KG}({\bm r},t)\nonumber\\
&=(x+iy)^l\frac{mK_{l+1}(ms)}{s^{l+1}}.
\end{align}
The wave packet in the form of the Macdonald function (in the simplest case of a massive spinless particle not carrying angular momentum) was obtained before by Naumov\&Naumov \cite{nn}.

Our prescription KG$\to$D works best in the Weyl (chiral) representation \cite{weyl} of $\gamma^\mu$ matrices,
\begin{align}\label{gamma}
\gamma^0=\left[\begin{array}{cc}0&I\\I&0\end{array}\right],\quad
\gamma^i=\left[\begin{array}{cc}0&-\sigma_i\\\sigma_i&0\end{array}\right],
\end{align}
where $I$ is the $2\times 2$ unit matrix. In this representation the Dirac equation $(i\gamma^\mu\partial_\mu-m)\Psi=0$ for the bispinor $\Psi=\{\phi,\chi\}$ splits into two coupled equations for two {\em relativistic} spinors,
\begin{align}\label{two}
i\sigma^\mu\partial_\mu\phi=m\chi,\quad
i\tilde{\sigma}^\mu\partial_\mu\chi=m\phi,
\end{align}
where $\sigma^\mu=\{I,\sigma_i\}$,  $\tilde{\sigma}^\mu=\{I,-\sigma_i\}$.
The Dirac equation in the Dirac representation of $\gamma^\mu$ matrices does not have this property; upper and lower components do not transform independently under Lorentz transformations.
Using the prescription KG$\to$D we construct the following four bispinors:
\begin{subequations}
\begin{align}\label{bsa}
\Psi_+&=N_l\left[\begin{array}{c}f_l\\0\\
i/m\,(\partial_t+\partial_z)f_l\\
i/m\,(\partial_x+i\partial_y)f_l
\end{array}\right],\\
\Psi_-&=N_l\left[\begin{array}{c}
i/m\,(\partial_t-\partial_z)f_l\\
-i/m\,(\partial_x+i\partial_y)f_l\\
f_l\\0\\
\end{array}\right],
\end{align}
\end{subequations}
\begin{subequations}
\begin{align}\label{bsb}
\Phi_+&=N_{l+1}\left[\begin{array}{c}0\\f_{l+1}\\
i/m\,(\partial_x-i\partial_y)f_{l+1}\\
i/m\,(\partial_t-\partial_z)f_{l+1}
\end{array}\right],\\
\Phi_-&=N_{l+1}\left[\begin{array}{c}
-i/m\,(\partial_x-i\partial_y)f_{l+1}\\
i/m\,(\partial_t+\partial_z)f_{l+1}\\
0\\f_{l+1}\\
\end{array}\right],
\end{align}
\end{subequations}
\noindent where $N_l$ are the normalization constants. All four bispinors $\Psi_\pm$ and $\Phi_\pm$ describe the states of a twisted electron: they are eigenstates of the $z$ component of the total angular momentum with the eigenvalue $(l+1/2)\hbar$. The solutions with negative eigenvalues can be obtained by the $180^\circ$ rotation around the $x$ axis.

In what follows we shall restrict ourselves to the bispinors $\Psi_\pm$ because the corresponding formulas for the bispinors $\Phi_\pm$ are more complicated. The derivatives in the definition of $\Psi_\pm$ can be easily evaluated leading to the expressions:
\begin{align}\label{bs1}
\Psi_+=N_l\left[\begin{array}{c}f_l\\0\\
\!\frac{a+it-iz}{x+iy}f_{l+1}\!\\
-if_{l+1}
\end{array}\right]\!,
\Psi_-=N_l\left[\begin{array}{c}
\!\frac{a+it+iz}{x+iy}f_{l+1}\!\\if_{l+1}\\
f_l\\0\\
\end{array}\right]\!.
\end{align}
The probability currents $j^\mu_\pm=\bar{\Psi}_\pm\gamma^\mu\Psi_\pm$ have the form
\begin{align}\label{cur1}
j^\mu_\pm=N_l^2\left[\begin{array}{c}|f_l|^2+\left(1+\frac{a^2+(t\mp z)^2}{x^2+y^2}\right)|f_{l+1}|^2\vspace{0.2cm}\\
\vspace{0.2cm}
2\frac{x(t\mp z)-ay}{x^2+y^2}|f_{l+1}|^2\\
2\frac{y(t\mp z)+ax}{x^2+y^2}|f_{l+1}|^2\vspace{0.2cm}\\
\pm |f_l|^2\pm\left(1-\frac{a^2+(t\mp z)^2}{x^2+y^2}\right)|f_{l+1}|^2\end{array}\right].
\end{align}

The normalization coefficients $N_l$ are defined by the condition $\int\!d^3r\,j^0=1$. For the bispinors $\Psi_\pm$ the integral of the probability densities $j_\pm^0=\Psi_\pm^\dagger\Psi_\pm$ evaluated at $t=0$ in the momentum representation is:
\begin{align}\label{norm}
&\int\!d^3r\,j_\pm^0=\frac{\pi N_l^2}{m^{2l+2}}\int\!d^3p\,
(p_x^2+p_y^2)^le^{-2aE_p}\nonumber\\
&=\frac{2\pi^{5/2} N_l^2}{m^{2l+2}}\frac{l!}{\Gamma(l+3/2)} \int_0^\infty\!dp\,p^{2l+2}e^{-2a\sqrt{m^2+p^2}}.
\end{align}
The integration over $p$ gives:
\begin{align}\label{norms}
N_l^{-2}=2m\pi^2 l!\,K_{l+2}(2am)/(am)^{l+1}.
\end{align}
For the bispinors $\Phi_\pm$ one must replace $l$ by $l+1$. As an illustration of the intricate topological properties of our solutions of the Dirac equation we plotted in Fig.~1 the integral lines of the current ${\bm j}_+$, i.e. the solutions of the set of three differential equations,
\begin{align}\label{intln}
\frac{d}{d\lambda}{\bm r}(\lambda)={\bm j}_+({\bm r}(\lambda),0).
\end{align}

\begin{figure}
\begin{center}
\vspace{-1cm}
\includegraphics[width=0.5\textwidth]{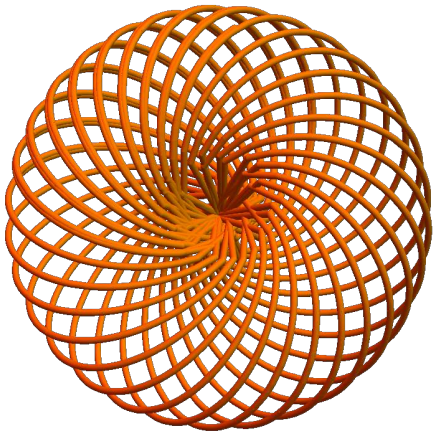}
\vspace{-1.5cm}
\caption{The linked lines of the current ${\bm j}_+$ for the simple Dirac hopfion ($l=1$), plotted at $t=0$.}\label{fig1}
\end{center}
\end{figure}

\section{Dirac hopfions vs. Maxwell hopfions}

The correspondence between the solution \eqref{kg} of the KG equation which generates the solutions of the Dirac equation and its massless counterpart \eqref{synge} generating the hopfion solution of the Maxwell equations serves already as a clear indication of a close relationship. In order to further justify the name hopfions for the states of electrons we choose the property of the solutions of the Dirac and Maxwell equations that can be directly compared. This will be the velocity field $\bm v$. In the case of electrons it is the current divided by the charge density: $\bm{v}_D=\bm{j}/j^0$ while in the case of the electromagnetic field it is the Poynting vector divided by the energy density: $\bm{v}_M=\bm{P}/E$. In order to underscore the close relation between the Dirac and Maxwell case, we shall write the velocity field for the bispinor $\Psi_+$ in the form:
\begin{align}\label{vd}
\bm{v}_D=\frac{\bm{w}+Q_l\bm{n}}{a^2+r^2+t^2-2tz+Q_l},
\end{align}
where $\bm{n}$ is the unit vector in the $z$ direction and
\begin{align}\label{q}
\bm{w}&=\left[\begin{array}{c}
2x(t-z)-2ay)\\
2y(t-z)+2ax)\\
r^2-a^2-t^2+2z(t-z)
\end{array}\right],\;
Q_l=\left|\frac{sK_{l+1}(ms)}{K_{l+2}(ms)}\right|^2.
\end{align}

\begin{figure}
\begin{center}
\includegraphics[width=0.5\textwidth]{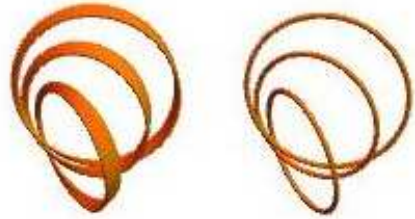}
\vspace{-2cm}
\caption{The linked lines of the velocity field ${\bm v}_+$ for the Dirac hopfion (left) and the Maxwell hopfion (right). The rings have very similar shapes. However, the Dirac rings do not close but keep winding.}\label{fig2}
\end{center}
\end{figure}

\begin{figure}
\vspace{-0.5cm}
\includegraphics[width=0.5\textwidth]{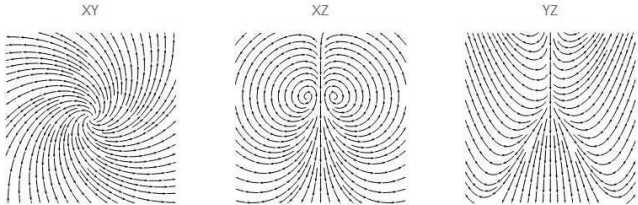}
\vspace{-0.5cm}
\caption{The streamlines of the Dirac hopfion velocity field built from the following components: $(v_x,v_y)$ on the $z=0$ plane, $(v_x,v_z)$ on the $y=0$ plane, and $(v_y,v_z)$ on the $y=0$ plane}\label{fig3}
\end{figure}

\begin{figure}
\vspace{-0.5cm}
\includegraphics[width=0.5\textwidth]{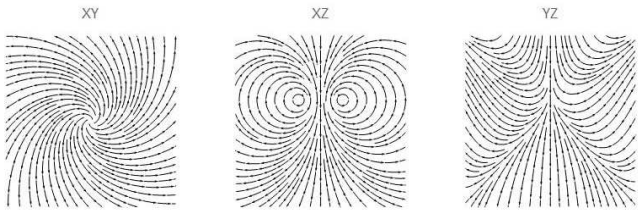}
\vspace{-0.5cm}
\caption{The streamlines of the Maxwell hopfion velocity field built from the components chosen in the same way as  in Fig.~\ref{fig3}.}\label{fig4}
\end{figure}

In the construction of the Maxwell hopfion from a scalar solution of the d'Alembert equation, there is an arbitrariness in choosing the phase of the polarization vector. We make here the same choice as in \cite{bb1}. The hopfion solutions of the Maxwell equations is generated by differentiation, as in \cite{bb1}, from any solution of the d'Alembert equation $g_l$. Choosing $g_l$ as a direct counterpart of the formula (\ref{arb}),
\begin{align}\label{g}
g_l=\frac{(x+iy)^l}{s^{l+2}},
\end{align}
we obtain the following Riemann-Silberstein vector,
\begin{align}\label{rs}
\bm{F}=\frac{\bm{E}+i\bm{B}}{\sqrt{2}}
=\frac{x_+^l}{s^{2l+6}}\left[\begin{array}{c}\vspace{0.2cm}
t_+^2-x_+^2\\
\vspace{0.2cm}
i(t_+^2+x_+^2)\\
-2t_+x_+
\end{array}
\right],
\end{align}
where $t_+=t+z-ia$ and $x_+=x+iy$. This is also an eigenstate of the total angular momentum, like our solutions of the Dirac equation; the eigenvalue is equal to $\hbar(l+1)$. The velocity field corresponding to this solution does not depend on the angular momentum,
\begin{align}\label{vm}
\bm{v}_M=\frac{\bm{w}}{a^2+r^2+t^2-2zt}.
\end{align}
The same vector field is obtained for the Dirac hopfion in the massless limit because when $m\to 0$ then $Q_l\to 0$. The qualitative features of the velocity fields $\bm{v}_D$ and $\bm{v}_M$ are still very similar even when $Q_l$ cannot be totally neglected. In Fig.\ref{fig2} we show the lines of the velocity vector $\bm{v}$ for the solutions of the Dirac equation and the Maxwell equations. In Figs.~\ref{fig3} and \ref{fig4}, we illustrate the similarities between these solutions by plotting the streamlines of the vector fields $\bm{v}_D$ and $\bm{v}_M$. The final argument that these velocity fields are intimately related comes from the comparison between the complex functions $\Upsilon=(v_x+iv_y)/(1-v_z)$. This function is often used in connection with Hopf fibering and it is exactly the same in the two cases:.
\begin{align}\label{ups}
\Upsilon_D=\Upsilon_M=\frac{x+iy}{t-z-ia}.
\end{align}
The level curves for this function are straight lines. For a given value of $\Upsilon$ the line is defined by the equations: $x=(t-z)\Re(\Upsilon)+a\Im(\Upsilon)$ and $y=(t-z)\Im(\Upsilon)-a\Re(\Upsilon)$. By stereographic projections, these functions become mappings of circles on the 3D sphere onto points on the 2D sphere: the standard Hopf fibration.

Considering the connections with the Hopf fibration and many similarities with the Maxwell hopfion, the term hopfion-like for our solutions of the Dirac equation seems to be well justified.

\section{Localization and spreading in time of hopfion wave packets}

The size of the wave packets for all freely moving particles changes in time in the same way: the mean square radius $<r^2>$ is a quadratic function of time. Owing to the time-reversal invariance, the evolution is symmetric in time $<r^2>=A+Bt^2$. The wave packets shrink until they reach their minimal extension and then they expand. For all Maxwell wave packets the rate is given by the exact formula $<r^2>_{\rm M}=A+(ct)^2$ and for the Maxwell hopfions $A=a^2$. For the Dirac hopfions the rate is given by the formula:
\begin{align}\label{spread}
<r^2>_{\rm D}=\lambdabar^2A_l(a/\lambdabar)+B_l(a/\lambdabar)(a^2+(ct)^2),
\end{align}
where $\lambdabar$ is the Compton wave length. The functions $A_l(a/\lambdabar)$ and $B_l(a/\lambdabar)$ can be expressed in terms of regularized hypergeometric functions $_pF_q(a/\lambdabar)$ or determined directly by numerical integration. The coefficient $B_l(a/\lambdabar)$ measures the rate of the wave packet spreading with time. In Fig.~5 we show the decrease of this rate with the increase of $a$ for several values of $l$. They all tend to 0.
\begin{figure}
\includegraphics[width=0.48\textwidth]{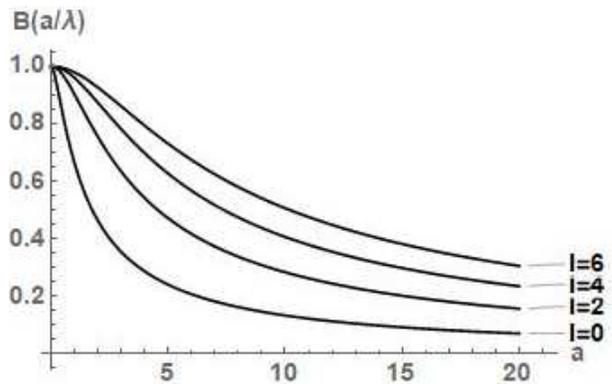}
\caption{The dependence of the rate of the spreading (in units of $c^2$) on the size of the Dirac wave packets for several values of $l$. The parameter $a$ is measured in Compton wave length.}\label{fig5}
\end{figure}
\begin{figure}
\centering
\includegraphics[width=0.48\textwidth]{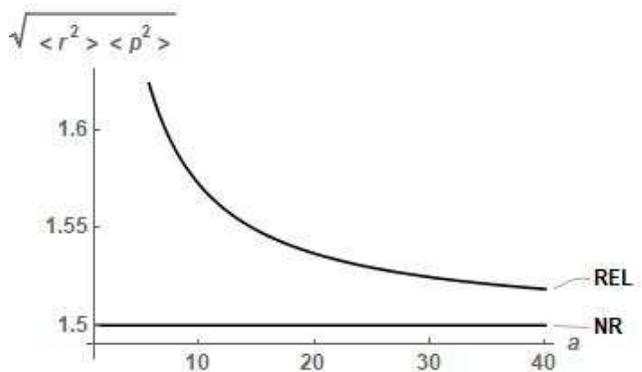}
\caption{The product of the uncertainties in position and momentum ( in units of $\hbar$) for the Dirac hopfion as a function of the size parameter $a$ measured in Compton wave length. In the limit, when $a\to\infty$, this product approaches the standard nonrelativistic limit $3/2\,\hbar$.}\label{fig6}
\end{figure}

When it comes to the Heisenberg uncertainty relations, in contrast to the Maxwell hopfion, the wave packet of the Dirac hopfion is much closer to the wave packets of the nonrelativistic quantum mechanics \cite{bb}. In Fig.~6 we show that the product of uncertainties in position and momentum for electrons approaches the nonrelativistic Heisenberg lower bound $3/2\,\hbar$, while for photons \cite{bb2} the sharp limit is $3/2\,\hbar\sqrt{1+4\sqrt{5}/9}$.

\section{Hopfions in motion}

So far we have considered Dirac hopfions in their rest frame. However, owing to the explicit relativistic invariance of our construction, we may easily put these hopfions in motion. The simplest method to find the effects of motion is by starting with the moving solution of the KG equation $f_l^v$ and generating from it the solution of the Dirac equation. In the coordinate system in which the hopfion is moving with velocity $v$ in the $z$ direction, we have:
\begin{align}\label{arbv}
f_l^v({\bm r},t)=(x+iy)^l\frac{mK_{l+1}(ms_v)}{s_v^{l+1}},
\end{align}
where $s_v=\sqrt{a^2+r^2-t^2-2ai(t-vz)/\sqrt{1-v^2}}$. The bispinors calculated from $f_l^v$ according the formulas (\ref{bsa}) and (\ref{bsb}) describe the states of electrons moving like spinning bullets in the $z$ direction. As was to be expected, their charge distribution is relativistically contracted. In Fig.~7 we show the shape of the moving hopfion.
\begin{figure}
\vspace{0.4cm}
\includegraphics[width=0.4\textwidth]{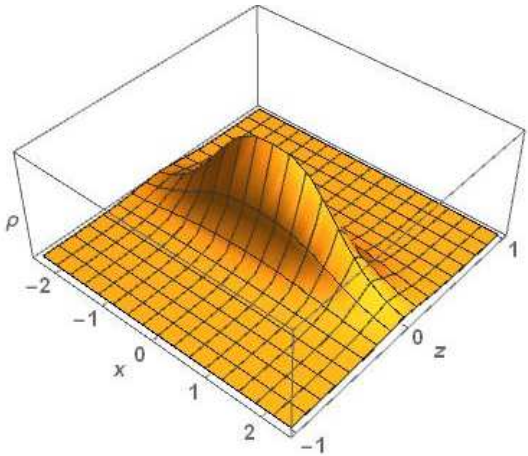}
\caption{The charge distribution of the hopfion ($l=0$) moving with the velocity $v=0.99c$ in the $z$ direction. The distribution has rotational symmetry around the $z$ axis but in the plot the $y$ direction is suppressed. The scale is in Compton wave length and $a=1$}\label{fig7}
\end{figure}

Our exact solutions of the Dirac equation describe relativistic electrons moving in free space but owing to their arbitrarily small size controlled by the parameter $a$ we can also describe their motion in slowly varying fields.

Topological properties of our hopfion-like solution of the Dirac equation are perhaps not as clear as those of the Maxwell hopfion where the Hopf fibration appears in its full form. However, considering the number of publications in many areas of physics, from nematic liquid crystals to gravitational waves, that employ hopfion-related concepts we hope that our solutions of the Dirac equation represent a viable addition to the hopfion zoo.

Numerical calculations and all figures were done with the use of Mathematica \cite{math}.

\end{document}